\documentstyle[12pt,aaspp4,flushrt,tighten]{article}

\begin{document}
%\twocolumn

\title{Mass Segregation in Young LMC Clusters I. NGC 2157
\footnote{Based on observations with the NASA/ESA {\it Hubble Space 
Telescope}
obtained at the Space Telescope Science Institute, which is operated by the
Association of Universities for Research in Astronomy Inc., under NASA
contract NAS 5-26555.}}

\author{Philippe Fischer\footnote{Hubble Fellow}\footnote{Visiting Astronomer,
National Optical Astronomy Observatories, which is operated by the Association
of Universities for Research in Astronomy, Inc., under contract to the National
Science Foundation.}}
\affil{Dept. of Astronomy, University of Michigan, Ann Arbor, MI 48109}
\authoremail{philf@astro.lsa.umich.edu}
\author{Carlton Pryor}
\affil{Dept. of Physics \& Astronomy, Rutgers, the
State University of New Jersey, P.O.\ Box 849, Piscataway, NJ 08855-0849}
\authoremail{pryor@physics.rutgers.edu}

\author{Stephen Murray}
\affil{Lawrence Livermore National Laboratory, L-023, P.O. Box 808, Livermore,
CA  94550}
\authoremail{murray8@llnl.gov}

\author{Mario Mateo}
\affil{Dept. of Astronomy, University of Michigan, Ann Arbor, MI 48109}
\authoremail{mateo@astro.lsa.umich.edu}

\author{Tom Richtler}
\affil{Sternwarte der Universit\"at Bonn, Auf dem H\"ugel 71,D-53121 Bonn}
\authoremail{richtler@astro.uni-bonn.de}

\abstract

We have carried out WFPC2 $V$- and $I$-band imaging of the young LMC cluster
NGC 2157. Construction of a color-magnitude diagram and isochrone fitting
yields an age of $\tau = 10^8$ yrs, a reddening $E(B-V) = 0.1$ and a distance
modulus of 18.4 mag.  Our data covers the mass range $0.75~{\cal M}_\odot \le m
\le 5.1~{\cal M}_\odot$.  We find that the cluster mass function changes
significantly from the inner regions to the outer regions, becoming steeper
(larger number of low mass stars relative to high mass stars) at larger radii.

The age of NGC~2157 is comparable to its two-body relaxation timescale only
in the cluster core.  The observed steepening of the mass function at larger
radii is therefore most likely an initial condition of the cluster stars.
Such initial conditions are predicted in models of cluster star formation
in which dissipative processes act more strongly upon more massive stars.

\section{Introduction} 

While some progress has been made towards understanding the formation of
isolated stars, the formation of star clusters is still poorly understood, even
though the majority of stars probably do form in clusters.  Until we obtain a
better understanding of the physical processes of cluster formation, knowledge
of star formation and the spectral, photometric, and chemical evolution of
composite stellar systems will remain significantly incomplete.

Obtaining observational constraints for cluster formation theory has been
very difficult. Star-forming Galactic clusters are sparse and tend to lie in
heavily obscured regions.  Older open clusters have relatively few stars and
low surface densities; thus, studies are plagued by small number statistics
and field star contamination. The globular clusters are all older than
$10^{10}$ yrs; stellar and dynamical evolution have erased most information
pertaining to their formation.

In contrast, the massive, young Magellanic Cloud clusters are excellent
laboratories for addressing fundamental issues of star formation in cluster
environments. They are relatively unobscured, possess rich populations of
stars, have high surface densities, and exhibit a large range in main-sequence
stellar masses. Their masses and other properties make some of these clusters
closely resemble the expected appearance of young Galactic globular clusters,
suggesting that these two kinds of clusters formed by similar processes.
Because these clusters are younger than their relaxation times (as young as
$10^7$ yrs) and orbit in the relatively weak gravitational field of the LMC,
they have had little dynamical evolution since formation, and so they may be
used to infer the conditions present in globular clusters immediately following
star formation.

In particular, the lack of extensive two-body relaxation in the LMC clusters
makes them good tools for discriminating among different theories of cluster
star formation.  Several theories differ crucially in whether or not extensive
dissipation is present during the cluster star formation process and in the
role played by the dissipative process in affecting star formation.  The
different theories therefore make significantly different predictions for the
behavior of the stellar initial mass function as a function of cluster radius,
which may be tested by HST observations.

Previous ground-based studies of the outer regions of young LMC clusters
(\cite{ma88}, \cite{el89a}, \cite{lee90}, \cite{sa91}, see the review in
\cite{ma92}) revealed mass functions similar to a Salpeter function in the
outer regions, but since the inner regions could not be resolved, mass
segregation could not be investigated. More recently, studies of R136
(\cite{hu95}) and NGC~1818 (\cite{hu97}) with HST have found Salpeter mass
functions and little evidence of mass segregation.

To test the cluster formation models we are studying four young LMC clusters
using HST WFPC2 data.  In this paper we present results for the first cluster,
NGC 2157, and focus on the search for primordial mass segregation in order to
help distinguish among the current plethora of cluster formation models and
provide strong constraints for subsequent, more detailed, models.
Section~\ref{observations} briefly describes the WFPC2 observations and
Sec.~\ref{scmd} discusses the color-magnitude diagrams and isochrone fitting.
Section~\ref{lumfunc} presents the cluster luminosity function and the evidence
for radial variation in the mass function.  Whether this mass segregation is
primordial is discussed in Sec.~\ref{discus}.  Section~\ref{conclusion}
discusses the implications of these results for theoretical models of cluster
formation and describes some future plans.

\section{Observations}	\label{observations}

Images of NGC~2157 were taken through two filters with WFPC2 on 9~Dec.  1995.
Exposure times were $5 \times 300$s and $2\times10$s through F555W and
$4\times300$s, $1\times 187.5$s and $1\times10$s through F814W.  The 187.5s
exposure was scheduled for 300s but was aborted due to telescope problems.  The
cluster center is located in the PC frame.  Additional exposures were taken of
a field offset 26\arcsec\ east and 110\arcsec\ north from the cluster field.
These consisted of $5\times300$s exposures through F555W and $5\times300$s
exposures through F814W.  There was a single integer-pixel dither for each
sequence of five 300s exposures.

The images have the normal pipeline preprocessing and the long exposures were
combined using the Tukey biweight algorithm (\cite{an72}; see \cite{be90}\ for
an astronomical application), which employs a robust weighting scheme to
estimate the mean (the aborted exposure was scaled by 1.6).  This technique was
very effective at eliminating cosmic rays and no cosmic rays are seen in the
final combined images.
%The combined PC cluster F555W
%and F814W images are shown in Fig \ref{combined}.

\section{Photometry and Color-Magnitude Diagrams} \label{scmd}

We used the profile-fitting photometry package DAOPHOT~II (\cite{st87},
\cite{st90}) with variable PSF to obtain stellar photometry for both the
cluster and background fields.  The photometry from ALLSTAR was aperture
corrected to a 0.5\arcsec\ radius aperture and the zero points, color terms,
and CCD gains from \cite{ho95} were adopted.  We also used the 0.04 mag ramp in
the CCD $y$ directions, as suggested by \cite{ho95}, to correct for
charge-transfer-efficiency problems.  The WFPC2 passbands were transformed to
Johnson $V$ and $I$ passbands as specified in \cite{ho95} with a reddening
correction of E$(B-V)=0.1$ (see below).  The estimated uncertainties in the
photometric zero points are around 0.05 -- 0.10 mag. The reddening corrected
photometry data for the cluster fields are tabulated on the AAS CD-ROM Vol. ?
as Tables \ref{tablecpc} - \ref{tablecwf4}, and the background fields in Tables
\ref{tableopc} - \ref{tableowf4}.

\begin{figure}
\plotone{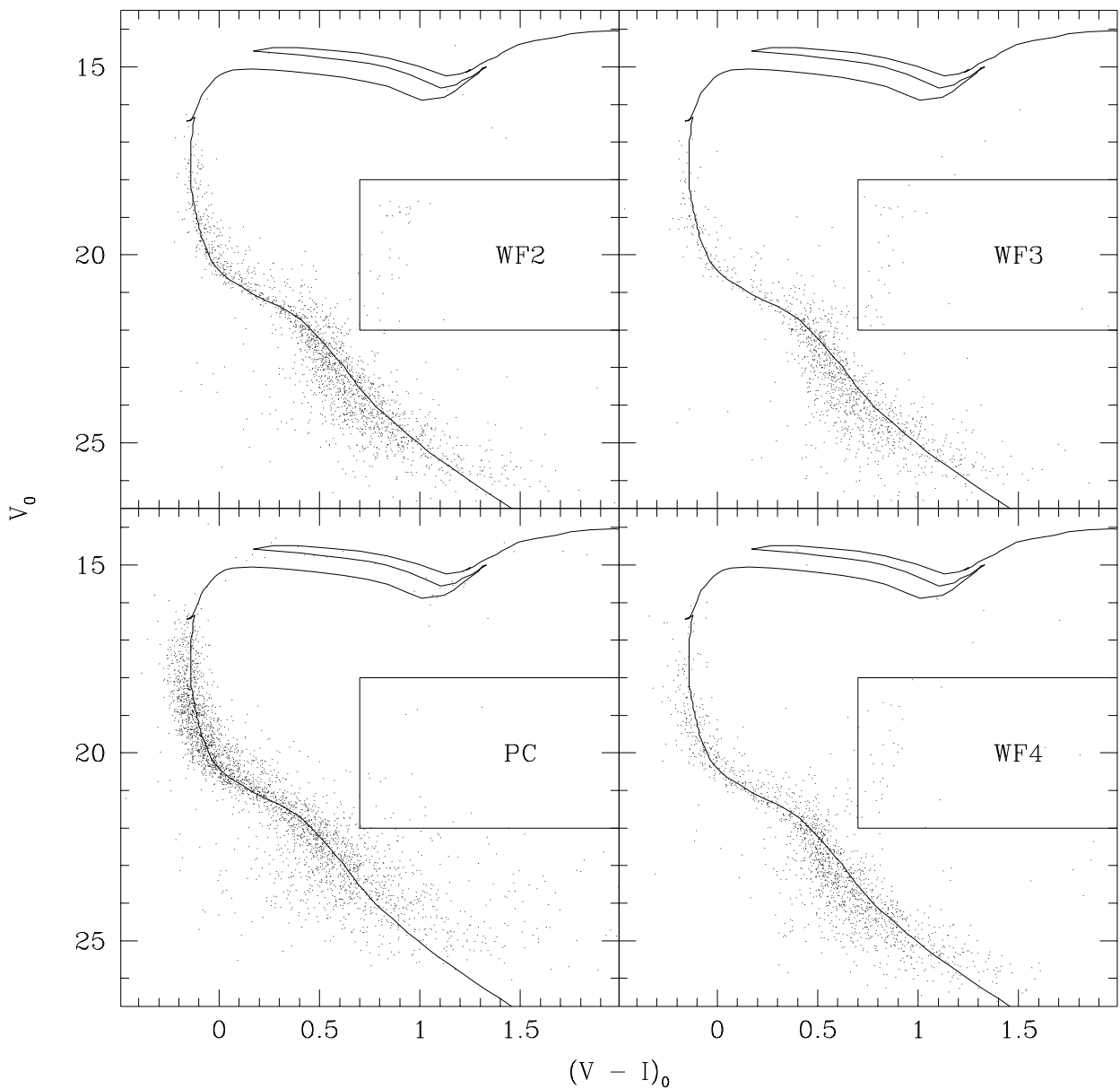}
\caption{$V$ and $I$ color-magnitude diagrams for the field around
NGC~2157.  The cluster center is located in the PC field.  The stellar
magnitudes have been dereddened assuming E$(B-V)=0.1$~mag.  The solid
line is the Bertelli et al. (1994) isochrone with $\tau = 1\times
10^8$~yrs and $Z = 0.008$, which fits the data well for the reddening
given above and $(m-M)=18.4$.  The boxes contain evolved old stars
which belong to the LMC field.  \label{cmd}}
\end{figure}

Figure~\ref{cmd} shows the $V_0$ {\it vs.} $(V-I)_0$ color magnitude diagrams
(CMDs) for all four WFPC2 fields.  The only stars displayed are those detected
independently in the two frames and whose positions, when transformed from one
frame to the other, agreed to within 1.0~pixel.  Because of saturation on the
long-exposure images, the short-exposure photometry was used for all stars with
$V_0<18.0$. 

The CMDs show clear evidence for a mixture of populations; a young population
belonging mainly to the cluster (with some LMC field stars mixed in) and an
older, evolved stellar sequence belonging exclusively to the LMC field.
Fitting $z=0.008$ isochrones from \cite{be94} to the young population yielded
E$(B-V)=0.1$~mag, $\tau = 1 \times 10^8$ yrs, and $(m-M)=18.4$.  The
metallicity of NGC 2157 is unknown at this time so we adopt z=0.008, which is
the approximate mean for the young LMC clusters.  We have dereddened the
apparent magnitudes using E$(B-V)=0.1$ and assuming A$_{F555W} = 0.3175$ and
A$_{F814} = 0.1895$ (Holtzmann et al. 1995).  This reddening is the same as
found by \cite{ma90} based on both $BVI$ photometry of NGC 2157 Cepheids and
comparisons of observed colors of upper main sequence stars with models of high
mass stars.  Errors in the photometric zero points lead to corresponding errors
in the distance modulus and reddening.  However, the age determination is less
sensitive to the photometric zero points since the age is determined by the
difference in brightness between the evolved stars and the strong main sequence
kink around $V = 21$.

\begin{figure}
\plotone{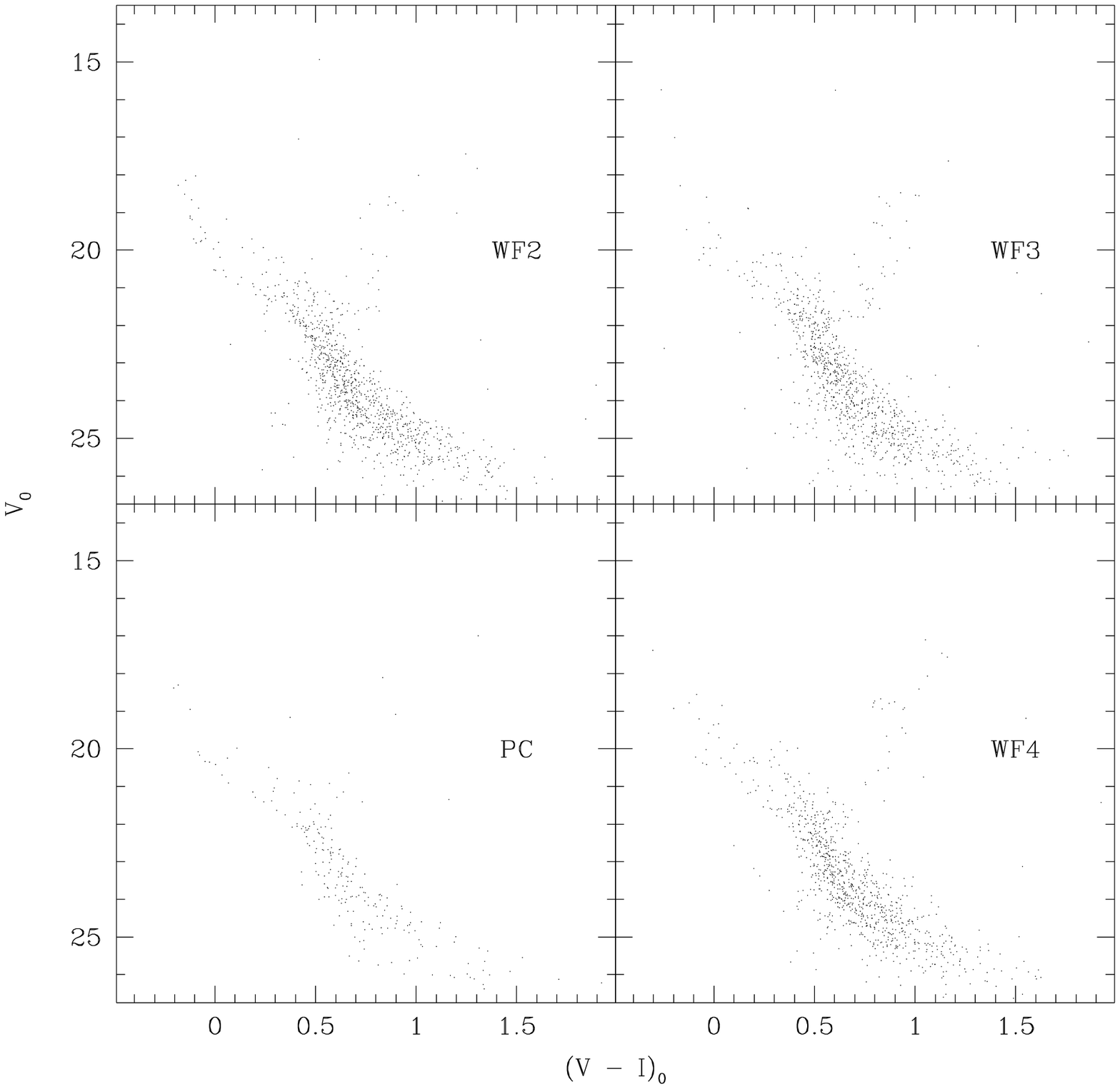}
\caption{$V$ and $I$ color-magnitude diagrams
for a field offset 110\arcsec\ from NGC~2157. The stellar magnitudes have been
dereddened assuming E$(B-V)=0.1$~mag. \label{cmdb}}
\end{figure}

Figure~\ref{cmdb} shows the CMDs of the background fields.  These also exhibit
a mixture of populations.

\section{Luminosity and Mass Functions} \label{lumfunc}

The main goal of this paper is to study the stellar mass function of NGC~2157.
The first step is to construct a luminosity function (LF hereafter) for the
cluster corrected for incompleteness and field star contamination.

\subsection{Completeness Corrections}

\begin{figure}
\plotone{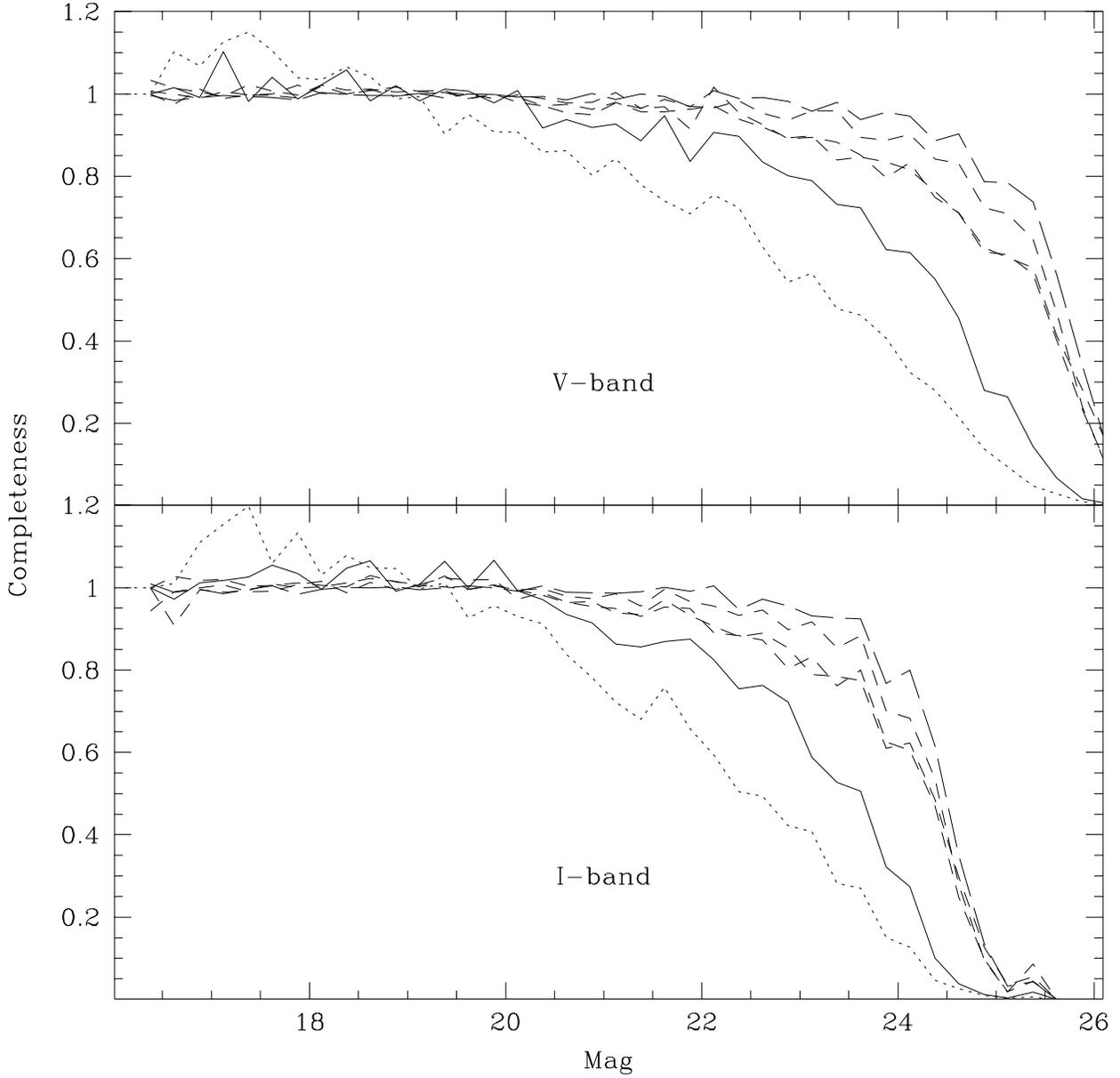}
\caption{Completeness fractions as a function of magnitude derived from
artificial star simulations for three different ranges of cluster radii. The
dotted line is the small-radius region ($R<11.8\arcsec; \bar{R} = 7.5\arcsec$),
the solid line is the intermediate-radius region (stars on the PC chip with
$R\ge12.3\arcsec$; $\bar{R}= 15.7\arcsec$), and the short-dashed lines are for
the three WF chips ($17.5 \le R \le 130$\arcsec; $\bar{R}=59.4$). The
long-dashed line show the average completeness curves for the three WF
background fields. Stellar magnitudes have been dereddened assuming
E$(B-V)=0.1$~mag
\label{complete}}
\end{figure}

We determined the incompleteness using the recovery of artificial stars added
simultaneously to our $V$ and $I$ frames.  The frames with added stars are
reduced in the same way as the originals.  Artificial stars are considered
detected only if they are independently found in both bandpasses, as were the
real stars.  We draw the stars randomly from a power-law mass function and
determine the $V$- and $I$-band brightnesses using the \cite{be94} isochrone
with $Z=0.008$ and $\tau = 1\times 10^8$~yrs.  Stellar positions are randomly
drawn from the radial distribution determined from stars with $V > 22.0$ for
each CCD image (a uniform distribution was used for the background fields).
The input mass function was varied until the recovered LF resembled the
measured LF (and hence the input mass function was similar to the cluster mass
function). By matching the recovered LF to the cluster LF we ensure that the
simulated data suffer from the same systematic effects as the real data. For
example, if some stars, due to blending, are measured too bright, then this
should be seen at a similar level in both the simulated and real data.  For
each WFPC2 image of both the cluster and background fields, we carried out a
minimum of 300 simulations, each containing 200 artificial stars.  In
Fig.~\ref{complete} we show completeness fraction curves for three different
cluster regions (described below) and the background fields.  The measured
completeness fraction is dependent on position; the regions near the cluster
center have substantially lower completeness fractions than regions at larger
radii due to crowding. The data were corrected by dividing by the completeness
functions in 0.25 magnitude bins.

\subsection{Luminosity Functions}

\begin{figure}
\plotone{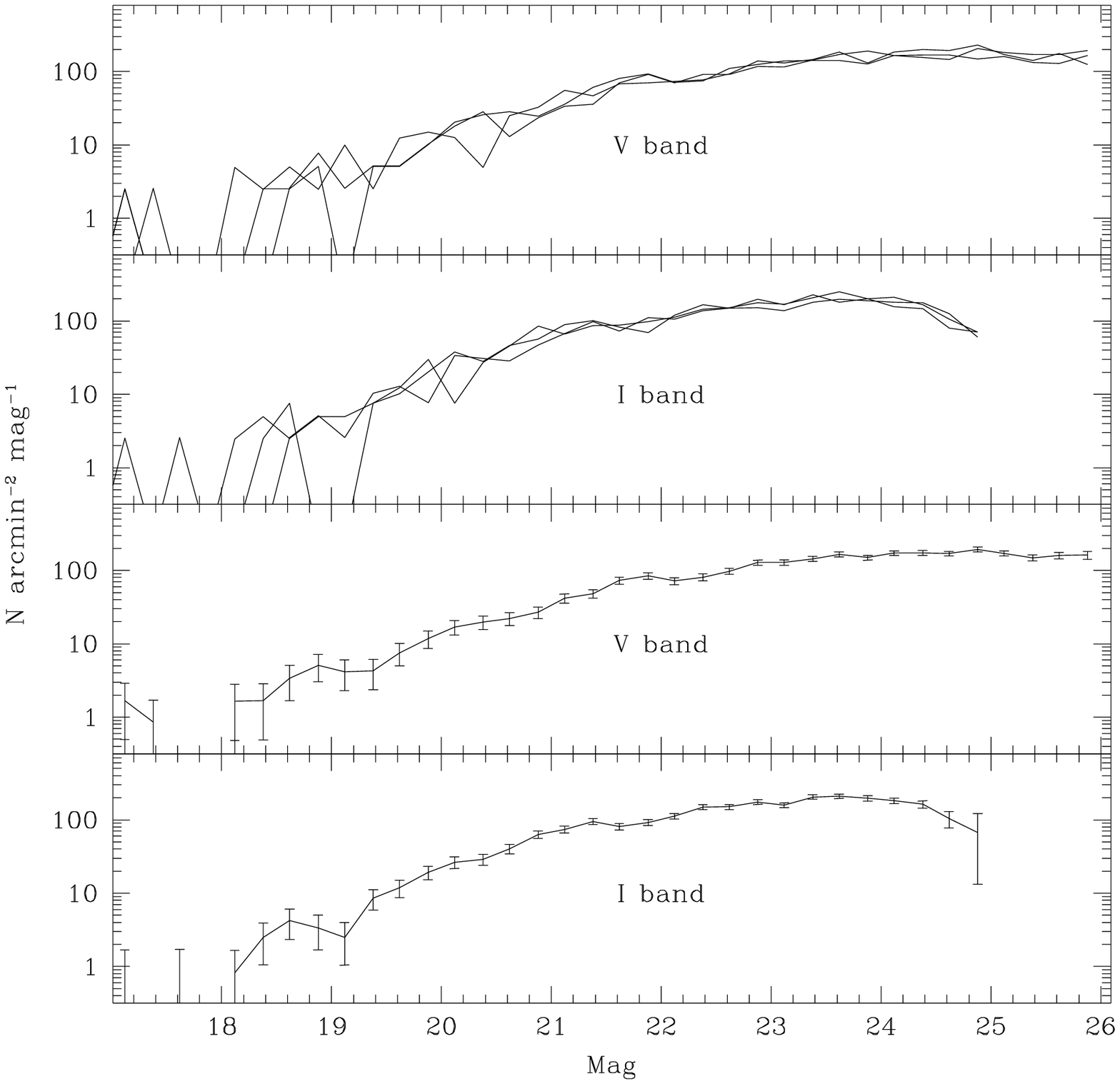}
\caption{The upper two panels show incompleteness-corrected, luminosity
functions for the three background WF images.  The bottom two panels show the
combined luminosity functions, where the error bars are based on
eq.~\protect\ref{uncereqn}.  The stellar magnitudes have been dereddened by
E$(B-V)=0.1$. \label{lumlo}}
\end{figure}

The upper two panels of Fig.~\ref{lumlo} show the completeness-corrected $V$-
and $I$-band LFs for the three WFC background fields.  Obvious evolved, old
stars (located in the box shown in Fig.~\ref{cmd}) have been excluded from all
background and cluster LFs shown in this paper.  The bottom two panels show the
{\it combined} $V$- and $I$-band LFs with $1\sigma$ error bars determined from
\begin{equation} \label{uncereqn}
\sigma \approx \sqrt{{n_{obs} \over f^2} + {(1-f)n^2_{obs} \over n_{added}f^3}}
\end{equation}
(\cite{bo89}), where $n_{obs}$ is the observed number of stars in the bin,
$n_{added}$ is the number of stars with magnitudes in the bin which were added
to the images for the artificial star simulations, and $f$ is the completeness
fraction of the bin from these simulations.  These analytical error bars seem
to agree reasonably well with the observed field-to-field scatter in the three
background LFs.

\begin{figure}
\plotone{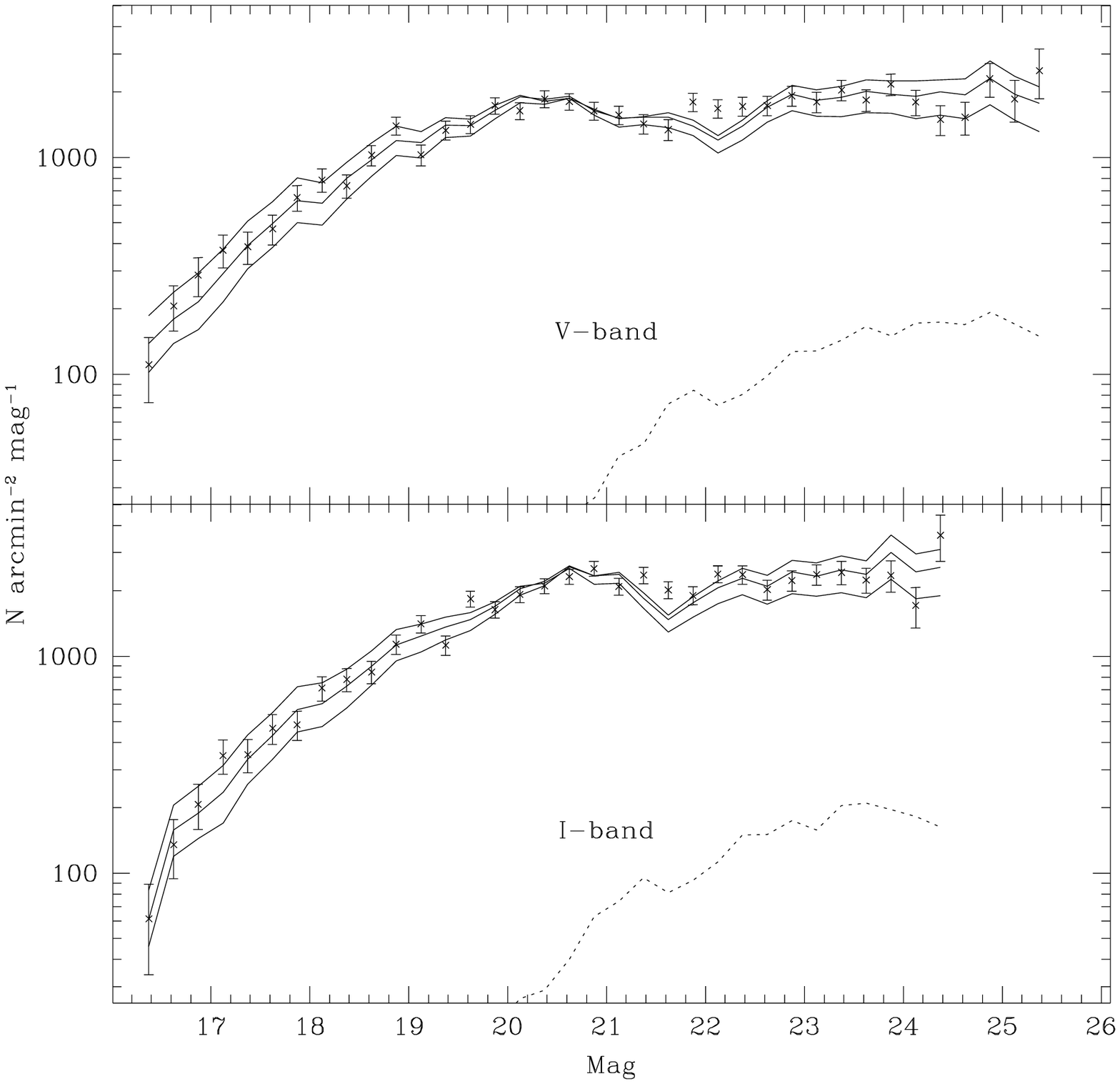}
\caption{$V$- and $I$-band, completeness-corrected, background-subtracted
luminosity functions for the PC field (which contains the cluster center). The
solid lines are synthetic luminosity functions corresponding to mass functions
having $x$ = 0.70, 1.00, and 1.25. The dashed lines show the background counts.
The stellar magnitudes have been dereddened by E$(B-V)=0.1$.\label{lumca}}
\end{figure}

\begin{figure}
\plotone{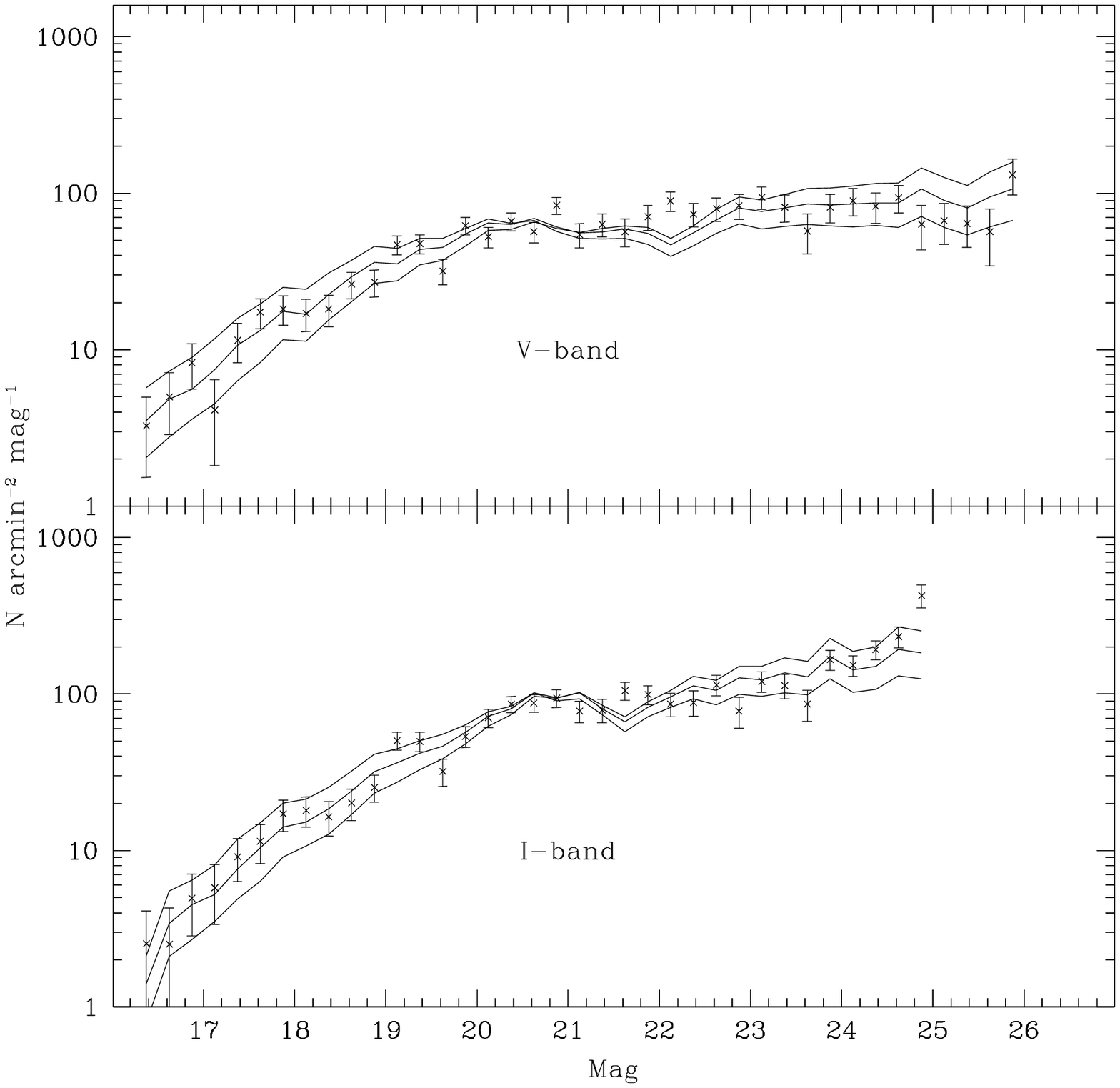}
\caption{Combined $V$- and $I$-band, completeness-corrected,
background-subtracted luminosity functions for the three WF frames.  The solid
lines are synthetic luminosity functions corresponding to mass functions with
$x$ = 0.85, 1.30, and 1.85 for the $V$-band data and $x$ = 1.10, 1.50, and 1.90
for the $I$-band data. The dashed lines show the background counts. The stellar
magnitudes have been dereddened by E$(B-V)=0.1$.  \label{lumwf}}
\end{figure}

Figure~\ref{lumca} shows the completeness-corrected, background-subtracted $V$-
and $I$-band LFs for the entire PC field, while Fig. \ref{lumwf} shows the
combined LFs for all three WFC chips.  Based on the \cite{be94} models, the
mass range represented by these LFs is approximately $0.75~{\cal M}_\odot \le m
\le 5.1~{\cal M}_\odot$.

To get a simple estimate of the mass function implied by our data, we assume
that it has the form of a single power-law:
\begin{equation}
{dN \over dm} \propto m^{-(1+x)},
\end{equation}
where $m$ is stellar mass, N is number of stars, and $x=1.3$ corresponds to the
Salpeter mass function. We convert this to a luminosity function using the
mass-luminosity relationship in \cite{be94} and compare it to the measured
luminosity function.  The power-law exponents that best fit the $V$- and
$I$-band LFs for the PC chip are $x_V = 0.95^{+0.30}_{-0.25}$ and $x_I =
1.0^{+0.25}_{-0.30}$, where the quoted uncertainties are the 95\% upper and
lower bounds. The reduced chi-square values for the fits are $\chi^2_{\nu} =
1.25$ ($\nu = 35$) and $\chi^2_{\nu} = 1.69$ ($\nu = 31$), respectively.  The
probability of seeing chi-square values that large or larger by chance is
P$(\chi^2) = 0.14$ for the $V$-band data and P$(\chi^2) = 0.0087$ for the
$I$-band data.  This is marginal evidence for the mass functions differing from
pure power laws. However, we have plotted synthetic LFs in Fig.~\ref{lumca} on
top of the measured LFs and the agreement appears good.  The largest
discrepancies appear around $V=22$ and $I=21.5$ and these look like a small
problem with the mass-luminosity relation. This slope is shallower than, but
consistent with, the mass function slope seen in the WFPC2 study of the young
LMC cluster NGC~1818 ($x=1.23 \pm 0.08$, \cite{hu97}).

It is worth briefly noting that the mass function slope derived here assumes
that the fraction of binary stars in the cluster is zero.  If the binary
fraction is non-zero, our mass function slope will tend to be too shallow
(\cite{sa91}). Lacking further information regarding the binary fraction, we
assume it is zero for the rest of this paper.

\subsection{Mass Segregation}

\begin{figure}
\plotone{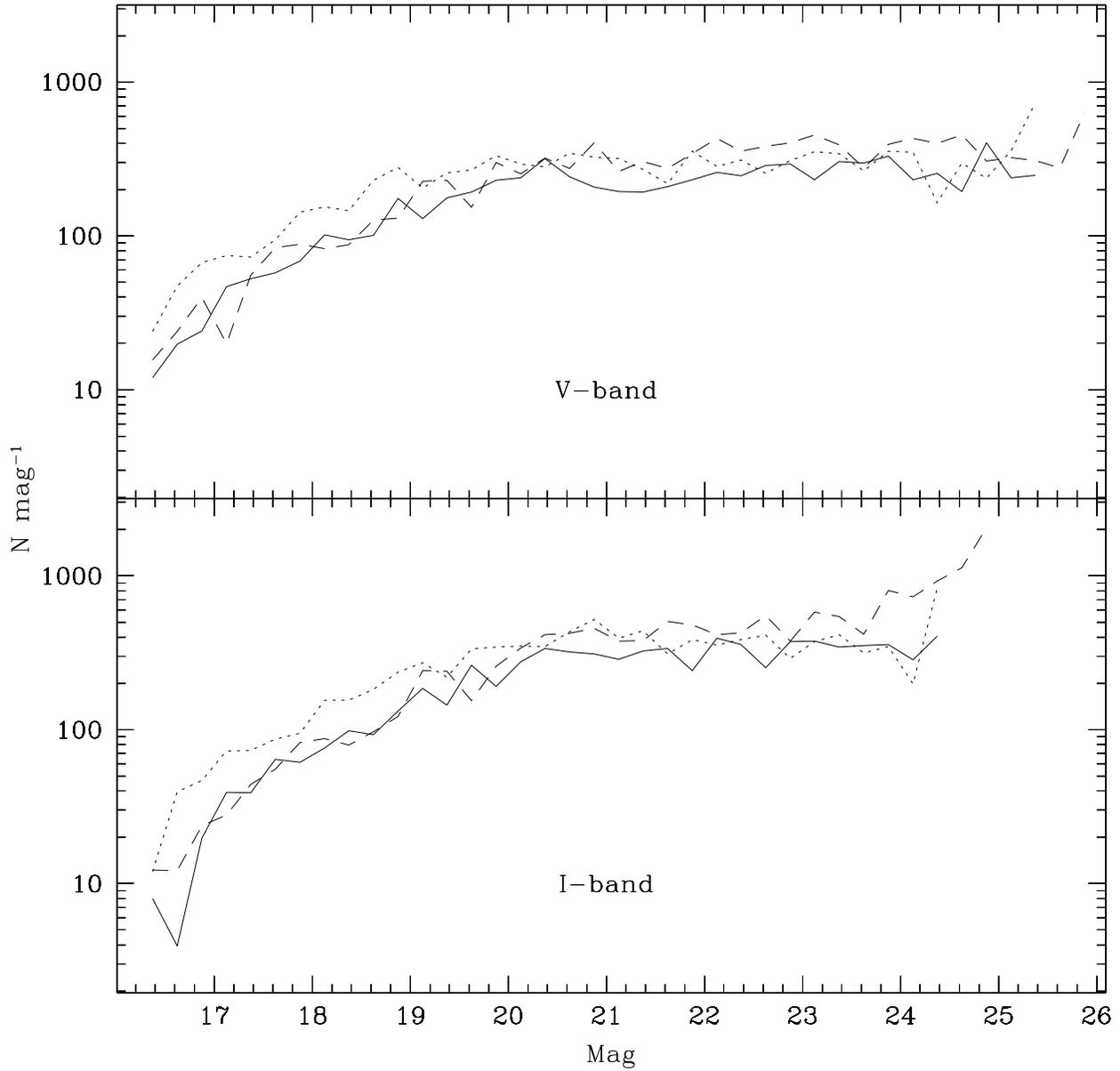}
\caption{Combined $V$- and $I$-band, completeness-corrected,
background-subtracted luminosity functions for the three radial regions
described in the text.  The dotted line is the small-radius region, the solid
line is the intermediate-radius region and the dashed line is the large-radius
region.  \label{lumcar}}
\end{figure}

To test for a variation of the mass function with projected distance from the
cluster center we divide our data set into three: a small-radius sample
($R<11.8\arcsec, \bar{R} = 7.5\arcsec$), an intermediate-radius sample (stars
on the PC chip with $R\ge12.3\arcsec$; $\bar{R}= 15.7\arcsec$), and a
large-radius sample (all of the stars on the WFC chips, $17.5\arcsec \le R \le
130$\arcsec; $\bar{R}=59.4\arcsec$).  Figure~\ref{lumcar} shows the $V$- and
$I$-band LFs for these three regions and they are tabulated in
Table~\ref{tablelf}. The best-fitting exponents for the different regions are
tabulated in Table~\ref{tablex}. If we remove the faintest (and most
discrepant) point from the $I$-band LF for the large-radius sample, we get the
same value for the mass function exponent, but the chi-square decreases to
$\chi^2_\nu = 1.52$.  There appears to be a trend towards steeper mass
functions with increasing radius. However, some of the fits to a power-law form
are poor, particularly at larger radii.  In order to quantify the significance
of the radial trend in the LF we carry out two model-independent comparisons.

First we employ a chi-square test between pairs of the LFs. One luminosity
function is optimally scaled (so as to minimize chi-square) and chi-square is
determined by summing the square of the differences in LFs divided by the
uncertainty over all the bins:

\begin{equation}
\chi^2_{ij} = \sum_{k=1}^n{{(s_{ij}b_{i,k} - b_{j,k})^2 \over
{s_{ij}^2\sigma_{i,k}^2+\sigma_{j,k}^2}}}.
\end{equation}

\noindent Here $b_{i,k}$ is the $k^{th}$ bin for region $i$, $\sigma_{i,k}$ is
the uncertainty in that bin, $n$ is the number of points and $s_{ij}$ is the
scaling factor which minimizes $\chi^2_{ij}$. The results of the $\chi^2$ tests
between the different regions are summarized in Table \ref{tablechi} for
different limiting magnitudes (which are given in columns 2 and 6). Columns 3
and 7 have the reduced $\chi^2$ and columns 4 and 8 are the degrees of
freedom. Columns 5 and 9 show the probability of exceeding the observed
$\chi^2$ value by chance.  The most significant differences occur between the
inner and outer regions for both bandpasses and all values of the limiting
magnitude. The differences between the inner and intermediate samples and the
intermediate and outer samples are not as significant.

The chi-square test does not fully take into account the information in the
systematic variation of the difference between two LFs from positive to
negative if they have different average exponents. Therefore the test will tend
to underestimate the significance of the differences in cases where the LFs
differ systematically. A Kolmogorov-Smirnov (KS hereafter) statistic applied to
the two cumulative LFs does use this information and should provide a more
sensitive test for differences between the LFs.  The difficulty with this
approach is that the corrections for incompleteness and the background require
that the confidence levels on this KS statistic be determined through Monte
Carlo simulations (see \cite{pa94}, for example).

For each of the radial samples we construct a normalized, background-corrected
cumulative luminosity function (CLF hereafter) that steps upwards by
\begin{equation}
\frac{f_i^{-1}}{\displaystyle\sum_{cl} f_i^{-1} - A\sum_{bkgd} f_j^{-1}}
\end {equation}
at the magnitude of each star in the radial sample of the cluster field and
steps downward by
\begin{equation}
\frac{Af_j^{-1}}{\displaystyle\sum_{cl}f_i^{-1} - A\sum_{bkgd} f_j^{-1}}
\end {equation}
at the magnitude of each star in the background field.  Here $f_i$ is the
completeness fraction at the magnitude of star $i$ in the specific radial
region of the cluster field, $f_j$ is the completeness fraction at the
magnitude of star $j$ in the background field, and $A$ is the area of the
radial region divided by the area of the background field.  The values of the
$f$'s used are those shown in Fig.~\ref{complete}.  The KS statistic is the
maximum vertical separation between the two CLFs.

\begin{figure}
\plotfiddle{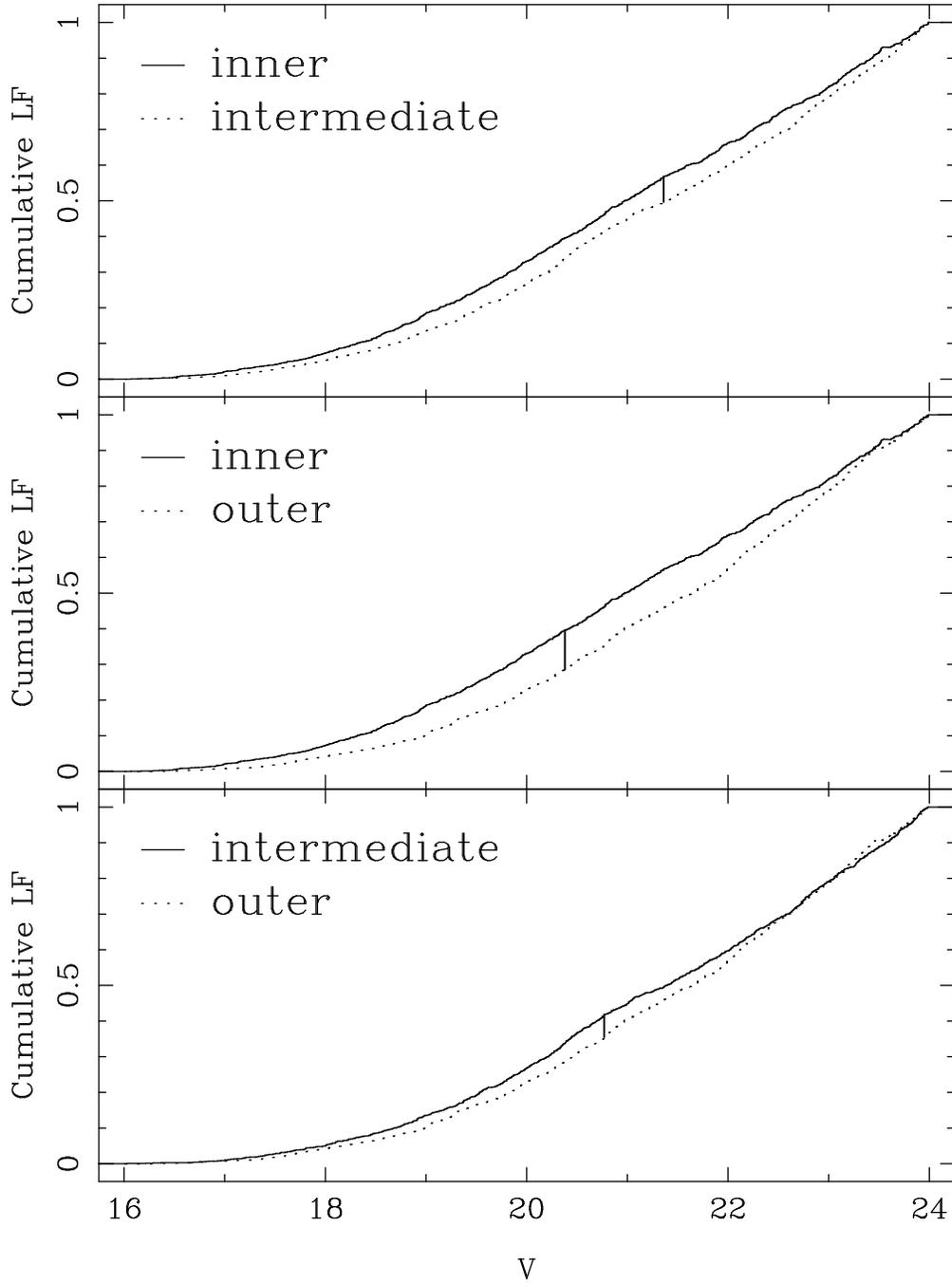}{6in}{0}{90}{90}{-275}{-100}
\caption{Comparisons between pairs of completeness-corrected,
background-subtracted, normalized cumulative luminosity functions for different
radial regions.  The vertical line marks the maximum vertical separation.
\label{ksfig}}
\end{figure}

The three panels of Fig.~\ref{ksfig} show the $V$-band CLFs for the three
pairings of the three radial regions.  The samples have been truncated at a
limiting magnitude of $V = 24$, where the inner, intermediate, and outer radial
samples are about 40\%, 60\%, and 80\% complete, respectively.  The vertical
lines in the figures mark the position of the largest vertical separation
between the two CLFs.  In all three panels, the CLF for the inner of the two
radial samples rises more quickly at bright magnitudes, signifying a shallower
luminosity function at smaller radii.  This is the case for every limiting
magnitude tried: $V$ = 25, 24, 23, and 22.  The $I$-band CLFs look similar to
those for the $V$-band.  Again, the shallower LFs occur at smaller radii for
limiting magnitudes of $I$ = 24, 23, 22, and 21.  The $I$-band LFs for the
inner, intermediate, and outer regions are about 40\%, 70\%, and 90\% complete
at $I = 23$.

We have performed Monte-Carlo simulations to determine if the differences seen
in Fig.~\ref{ksfig} are significant.  The average of the
incompleteness-corrected, background-subtracted LFs of the two regions,
weighting by the incompleteness-corrected number of stars, is the LF that we
use to test the null hypothesis that the two LFs are identical.  For each
sample, this LF is modified by incompleteness and the background is added to
produce ``observed LFs'' from which two samples of the observed size are drawn.
A background sample with the same size as the real background sample is drawn
from the real sample with replacement.  We calculate the KS statistic for these
artificial data and repeat the process 10,000 times.  If the number of times
that the statistic exceeds the value actually observed is small, the null
hypothesis can be rejected.

Table~\ref{kstable} gives the observed KS statistic and the fraction of the
simulations for which the statistic exceeded this value for the three pairings
of the radial regions.  The leftmost column gives the regions compared and is
followed by two blocks of seven columns giving numbers for the $V$-band and
$I$-band LFs.  These seven columns are the limiting magnitude, the numbers of
stars in the inner, outer, and background samples, the maximum vertical
separation between the two LFs, the magnitude at which this occurs, and the
fraction of the simulations with a larger maximum vertical separation than was
actually observed.  Our Monte-Carlo simulations do not include the effect of
the uncertainties in the incompleteness corrections, but we present the results
for a selection of limiting magnitudes.  For the brighter limits, corrections
for incompleteness have had little effect on the LFs.  The confidence with
which the hypothesis of identical LFs can be rejected varies little with
limiting magnitude, except to become somewhat less certain at the brightest
magnitudes.  This is expected because of the smaller sample sizes and the
shrinking range of luminosities.

Identical luminosity functions for the inner and outer radial regions can be
rejected at better than 99\% confidence for nearly all of the limiting
magnitudes.  Similarly, the LFs of the inner and intermediate regions are
different with better than (usually much better than) 95\% confidence.  The
difference between the intermediate and outer region LFs are not significant.

Our detection of mass segregation in NGC~2157 differs from what was found in
the $\tau = 20$~Myr LMC cluster NGC 1818 by Hunter et al.\ (1997).  Stars in
their PC field, centered on the cluster, yielded $x = 1.21 \pm 0.10$, while the
stars in their WFC fields yielded $x = 1.25 \pm 0.08$.  Similarly, the very
dense, $\tau = 3$~Myr LMC cluster R136 shows no mass segregation, except
perhaps within a radius of 0.4~pc (\cite{hu95}).  We thus turn to a discussion
of the possible cause of our observed mass segregation.

\section{Mass Segregation: Primordial or Evolutionary?}

\subsection{Evolution of Mass Segregation: Theory}

The first step in deciding whether the mass segregation described in the
previous section is primordial is to estimate the relaxation time at different
points in the cluster.  While theory and simulations show that complete energy
equipartition is unlikely to be established in a star cluster (\cite{sp69},
\cite{in85}, see the summary in \cite{me97}), mass segregation certainly does develop on the timescale
to exchange energy between stars of different mass by small-angle scattering,
$t_{eq}$.  If all of the stars in a cluster begin with the same spatial
distribution stars of mass $m_1$ will at least initially have $t_{eq,1} =
(\langle m \rangle/m_{1}) t_r$, where $t_r$ is the local two-body relaxation
time (\cite{sp69}).  Thus, mass segregation will be most rapid for the most
massive stars and where $t_r$ is shortest (at the smallest radii).

Numerical simulations with a range of stellar masses ({\it e.g.}  \cite{in85};
\cite{ch90}) do show the profiles of the heaviest stars changing quickest and
the effects of segregation propagating outward through the cluster.  If stars
with different masses start with the same velocity dispersion, the ``cooling''
of the heavy stars at small radii can leave a kink in their radial density
profile (see Fig.\ 2 of \cite{in85}, for example).  The simulations show that
significant mass segregation among the heaviest stars in the core occurs in the
local relaxation time, $t_{r0}$, but affecting a large fraction of the mass of
the cluster requires a time comparable to the average relaxation time averaged
over the inner half of the mass, $t_{rh}$.

\subsection{Relaxation Timescales for NGC 2157}

We will estimate the relaxation times in NGC~2157 from its current density
profile, but we point out that the profile may have been different in the past.
The rapid loss of mass due to the evolution of massive stars can cause a
cluster to expand (\cite{ag79}, \cite{ch90}).  Observational support for this
idea is provided by an apparent increase in the upper limit to the core radii
of LMC clusters with age (\cite{el89b}).  Thus, it is possible that NGC~2157
was denser and had shorter relaxation times when it formed than it does today.
However, this same expansion of the core acts to reduce, though not eliminate,
mass segregation by expanding the spatial distribution of the heavy stars in
the core.  Detailed simulations will probably eventually be required to
estimate how these processes have affected NGC~2157.

\cite{el91} gives a surface brightness profile for NGC~2157 derived from V and
B CCD images.  These yielded King-model core radii, $r_c$, of 9.6\arcsec $\pm$
1.3\arcsec\ and 9.2\arcsec\ $\pm$ 0.8\arcsec, respectively, which became
8.6\arcsec\ and 7.9\arcsec\ after being corrected for the seeing.

We performed surface photometry on a 400~s V-band image obtained with the CTIO
0.9~m telescope by PF on January 23, 1995.  The stellar images had a full width
at half-maximum of 2.0\arcsec.  We found the center of the cluster using
mirrored autocorrelation (\cite{dj87}, \cite{he85}) in a box with sides of
40\arcsec.
%To reduce the effect of individual bright stars, we actually used an image
%formed by taking the logarithm of the pixel values.  
Table~\ref{sbtable}\ gives the surface brightness profile calculated
% from the original image
in circular annuli about this center.  From left to right, the columns are the
base-10 logarithm of the radius in arcseconds, the surface brightness in V
magnitudes per square arcsecond, and the uncertainty in the surface brightness
calculated from the scatter between the octants.  The photometric calibration
is based on V magnitudes from
\cite{el91} for stars in the image.

The profile in Table~\ref{sbtable}\ resembles that of \cite{el91} in having a
``shoulder'' at a radius of about 5\arcsec, but this shoulder is less prominent
in our data, presumably because of a slightly different center.  A profile
calculated from the stars detected in the PC image (but not corrected for
incompleteness) shows no shoulder, suggesting that its presence in the surface
brightness profiles is due to noise.  This eliminates one potential piece of
evidence for a lack of relaxation in the core.

The single-component King (1966) model plus a background that best fits the
profile in Table~\ref{sbtable}\ has $r_c = 8.7$\arcsec~$\pm$~0.6\arcsec, a
central surface brightness of $\mu_{0,V} = 17.27 \pm 0.11$~mag~arcsec$^{-2}$, a
concentration of $c \equiv \log(r_c/r_t) = 1.1 \pm 0.1$, and a background of
$20.16 \pm 0.005$~mag~arcsec$^{-2}$.  This core radius is very similar to that
found by \cite{el91}, so we adopt $r_c = 8.7$\arcsec\ for the remainder of this
paper.  (This ``core radius'' is where the surface brightness falls to half of
its central value.  The King-model scale radius is about 10\% larger.)

To estimate relaxation times we also need a mass for NGC~2157.  One way to
proceed is to combine our luminosity functions with the above King model fit.
The luminosity function for the innermost region from Fig.~\ref{lumcar} implies
a total of 2280 stars with $r < 11.8$\arcsec\ and masses between 0.75~${\cal
M}_\odot$ and 5.1~${\cal M}_\odot$.  With our fitted value for the core radius
% this number
% implies a central surface density of $\Sigma_0 =15$~stars~arcsec$^{-2}$.
% For a King model with $c=1.1$, $\Sigma_0 = 1.8 n_0 r_c$, where $n_0$ is
% the central density and the coefficient varies only slowly with
%concentration.  For 
and our adopted distance of 48~kpc to NGC~2157, $r_c = 2.0$~pc and the central
density of stars with masses between 0.75~${\cal M}_\odot$ and 5.1~${\cal
M}_\odot$ is $n_0(0.75,5.1) = 78$~stars~pc$^{-3}$.

If we adopt a mass function power-law exponent of 1.0, which is close to the
average for all of our data, then the central mass density of observed stars is
$\rho_0(0.75,5.1) = 130~{\cal M}_\odot$~pc$^{-3}$.  Assuming that stars with
initial masses between 5.1 and 8~${\cal M}_\odot$ form 1.0~${\cal M}_\odot$
white dwarfs and that stars between 8 and 80~${\cal M}_\odot$ form 1.4~${\cal
M}_\odot$ neutron stars (all of which are retained in the cluster) adds only
another 16~${\cal M}_\odot$~pc$^{-3}$ to the central density if the same
exponent held above the present-day turnoff.  Now there must be at least some
stars fainter than we observe.  If we extrapolate the same power-law mass
function to lower masses, then the density in main sequence stars is, for
example, $\rho_0(0.1,5.1) = 270~{\cal M}_\odot$~pc$^{-3}$ or $\rho_0(0.03,5.1)
= 350~{\cal M}_\odot$~pc$^{-3}$.

If the lower cut-off of the mass function is 0.1~${\cal M}_\odot$, then $\rho_0
= 290~{\cal M}_\odot$~pc$^{-3}$ and the mean stellar mass is $\langle m \rangle
= 0.45~{\cal M}_\odot$.  With these numbers, the best-fitting King model
described above yields a total mass for the cluster of ${\cal M} = 4.4\times
10^{4}~{\cal M}_\odot$, a central relaxation time of $t_{r,0} = 9.1\times
10^{7}$~yrs, and $t_{rh} = 9.8\times 10^{8}$~yrs.  Over the range of King
models with acceptable fits to the surface brightness profile, $t_{r,0}$ varies
by $\times 4.6$ and $t_{rh}$ by $\times 1.5$.

Clearly, it is desirable to have some check on the extrapolation of the mass
function made above.  Observations with the Rutgers Fabry-Perot on the CTIO 4~m
telescope have yielded radial velocities for 70 stars in NGC~2157.  These data
will be presented in more detail elsewhere, however, we use them here to
provide a check on our relaxation times.  Fitting these data with the $c=1.1$
King model yields $\rho_0 = 470~{\cal M}_\odot$~pc$^{-3}$, ${\cal M} =
7.2\times 10^{4}~{\cal M}_\odot$, $t_{r,0} = 1.4\times 10^{8}$~yrs, and $t_{rh}
= 1.2\times 10^{9}$~yrs.  While the effect of the velocity measurement errors
on the dispersion are removed by our fitting procedure (see \cite{pr89} for
more details), the dispersion might still have been inflated by the orbital
motions of binary stars, for example.  The truth is likely somewhere between
the two sets of values.

\subsection{Interpretation}

Thus, the $1\times 10^8$~yr age of NGC~2157 is comparable to its $t_{r,0}$ and
about a tenth of its $t_{rh}$.  If these times have not been much shorter over
most of the age of the cluster, then we expect only a small amount of mass
segregation to have occurred in the cluster and it should not extend much
beyond the core.  If the initial mass function was a power law with exponent
1.0 and extended up to 80~${\cal M}_\odot$, then stellar evolution has caused
the cluster to lose about half of its mass.  This would lead to significant
cluster expansion, but the amount of mass loss is sensitive to the adopted
upper cutoff and the assumption of a constant exponent.  In our galaxy, the
initial mass function has an exponent of about 1.6 at large mass ({\it e.g.},
\cite{ba92}).  Also, much of the stellar mass loss would occur very early, so
any dense initial stage should be short-lived.  More definitive statements will
require detailed modeling of the dynamical evolution of the cluster, but we
conclude that it is unlikely that significant mass segregation has occurred
outside of the core since NGC~2157 formed.

\section{Discussion} \label{discus}

The nature and degree of primordial mass segregation in a star cluster is
determined by the importance of interactions during the star formation process
in clusters.  Different theories which have been proposed to explain cluster
star formation lead to very different predictions for primordial mass
segregation.

In the classic picture of star formation (\cite{sh87}), protostars evolve in
isolation.  Initial objects having masses in excess of $M_G$ (the critical mass
for gravitational collapse) form protostellar cores, which accrete gas until
infall is halted by winds, which are assumed to result from the onset of
deuterium burning.  Interactions among the protostellar cores are assumed to be
unimportant.  The initial evolution of the cluster therefore occurs by violent
relaxation and so no mass segregation is expected (\cite{lb67}).

The results are quite similar in the picture of Burkert et al. \markcite{bu93}
(1993).  In their model, an initial generation of massive stars enriches a
primordial gas cloud.  Stellar winds and supernovae lead to expansion of the
cloud, which then undergoes star formation, and, subsequently, violent
relaxation.

In other models of cluster star formation, interactions among the protostellar
cores play a crucial role in the star formation process.  The nature of the
mass segregation to be expected is determined by whether or not interactions
act to enhance or to halt the accretion process.

Podsiadlowski \& Price \markcite{po92} (1992) propose a picture in which the
initial stages of star formation closely resemble the picture of Shu et
al. \markcite{sh87} (1987).  In their model, however, accretion onto a
protostellar core is terminated when the infalling gas cloud is disrupted by a
close encounter with another protostellar core.  Primordial mass segregation in
this model results from the changing ratio of the timescales for protostellar
collisions and of gas infall onto the protostars.  For example, if the gas and
protostars in a cluster both follow isothermal density distributions as a
function of cluster radius, then

\begin{equation}
\tau_{coll}/\tau_{infall}\propto R,
\end{equation}

\noindent
where $R$ is the radius within the cluster.  The increase of the ratio with $R$
implies that accretion can proceed further for protostellar cores at larger
cluster radii, leading to the formation of more massive stars.  In this model,
therefore, primordial mass segregation is expected in the sense that the IMF
should become shallower (have more massive stars) at larger radii.

The exact reverse is expected if encounters {\it enhance} mass accretion.  In
the picture of Murray \& Lin \markcite{mu93}\markcite{mu96}(1993, 1996), the
fragmentation of a protocluster cloud is triggered by thermal followed by
dynamical instability.  Thermal instability is governed by local heating and
cooling balance, and so has no preferred scale length, while the dynamical
instabilities involved act preferentially upon smaller scale lengths.  The
result is therefore a cluster of non-self-gravitating cloudlets.  The cloudlets
cannot contract individually until their masses exceed $M_G$, but the cluster
as a whole does contract due to the mutual gravity of the cloudlets.  As the
cloud contracts, mergers among the cloudlets eventually lead to their masses
exceeding $M_G$, at which point they contract to form protostars.  Continued
encounters can increase the protostellar mass above $M_G$.  The mergers lead to
extensive dissipation of kinetic energy.  More massive stars undergo more
mergers, and so dissipate more energy.  They also tend to form near the cluster
center, where the density and, hence, the encounter rate are highest.
Primordial mass segregation is therefore expected in the sense that the IMF
should become steeper (contain a greater fraction of low mass stars) at larger
cluster radii.

This last scenario appears to be the most consistent with our observations.  We
have detected mass segregation in the young LMC cluster NGC 2157 at a high
level of confidence. It appears that the relative number of low to high mass
stars is an increasing function of radius.  Although the results of
observations of other clusters is needed to strengthen any conclusions, it is
of interest to compare the observations of NGC~2157 with the results of Murray
\& Lin (1996).  The mass function slopes found for NGC~2157 ($x\approx1$), and
the relatively weak change in the mass function with cluster radius are best
represented by models in which the initial mass of the cloudlets, $M$, is a
large fraction of $M_G$, and in which the initial ratio of the kinetic to
potential energy of the cluster $Q=0.01$ ($Q=0.5$ in a virialized cluster).  In
these models, only one or two mergers are required before the onset of
protostellar collapse.  The population of cloudlets is therefore depleted
relatively early in the collapse of the cluster, before the density reaches its
peak.  Fewer high mass stars are formed, and violent relaxation plays a greater
role than in models in which the initial cloudlet mass was a smaller fraction
of $M_G$, for which protostar formation was delayed until later in the
collapse.  More extensive mass segregation is also seen in models in which
$Q=0.5$.

\section{Conclusions and Future Work} \label{conclusion}

Based on WFPC2 imaging of the young LMC cluster NGC 2157 we have detected
evidence for mass segregation in the sense of a steeper mass function at larger
radius.  An analysis of the two-body relaxation times indicates that this mass
segregation is most likely mainly primordial and not evolutionary.  If it is
indeed primordial, then models of cluster formation in which encounters between
protostars enhance mass accretion (\cite{mu93}, \cite{mu96}) are favored while
models where encounters halt mass accretion are ruled out.

However, studies of two other young LMC clusters did not find evidence for mass
segregation (\cite{hu95}, \cite{hu97}).  We have additional WFPC2 data for two
other such clusters, NGC~2004 and NGC~2031, as well as scheduled observations
for NGC 1711.  We will carry out similar analyses of these clusters in order to
better understand our result for NGC~2157.  NGC~2004 and NGC~1711 will be
particularly interesting as they are even younger than NGC~2157 and therefore
evolutionary mass segregation has had less time to operate.

Support for this work was provided by NASA through grants \#HF-01069.01-94A and
\#GO-05904.01-94A from the Space Telescope Science Institute, which is operated
by the Association of Universities for Research in Astronomy Inc., under NASA
contract NAS5-26555. We thank Karl Gebhardt for analyzing the NGC~2157
Fabry-Perot data.

%Mass function - whole PC field = 
%	V = 0.95 (+0.30)(-0.25) 95% confidence.  chi^2=1.25 (nu=36)
%	I = 1.0 (+0.25)(-0.30)    ''		chi^2=1.69 (nu=32)
%	       - r < 260
%	V = 0.80 (+0.40)(-0.40) 95% conf	chi^2=1.37 (nu=36)
%	I = 0.80 (+0.35)(-0.35)   ''		chi^2=1.32 (nu=32)
%	       - r > 270
%	V = 1.10 (+0.45)(-0.40)   ''		chi^2=0.92 (nu=36)
%	I = 1.15 (+0.40)(-0.35)			chi^2=1.66 (nu=32)
%		- summed WF chips
%	V = 1.30 (+0.45)(-0.45)			chi^2=1.53 (nu=38)
%	I = 1.50 (+0.40)(-0.40)			chi^2=1.83 (nu=34)
%	remove faintest point
%	I = same				chi^2=1.55 (nu=33)
{}

\begin{table}
\dummytable\label{tablecpc}
\end{table}
\begin{table}
\dummytable\label{tablecwf2}
\end{table}
\begin{table}
\dummytable\label{tablecwf3}
\end{table}
\begin{table}
\dummytable\label{tablecwf4}
\end{table}
\begin{table}
\dummytable\label{tableopc}
\end{table}
\begin{table}
\dummytable\label{tableowf2}
\end{table}
\begin{table}
\dummytable\label{tableowf3}
\end{table}
\begin{table}
\dummytable\label{tableowf4}
\end{table}

%\begin{figure}
%Combined F555W ($5\times300$s) and F814W ($4\times300$s, $1\times187.5$s) PC
%images of the central region of the cluster. The field is 34\arcsec\ on a side.
%\end{figure}

\def\baselinestretch{0.96}

\begin{deluxetable}{ccccccc}
\tablewidth{0pt}
\tablecaption{Luminosity Functions}
\tablehead{
\colhead{} & \multicolumn{2}{c}{Inner Region} & \multicolumn{2}{c}{Intermediate Region} & \multicolumn{2}{c}{Outer Region} \\
\colhead{Mag\tablenotemark{1}} & \colhead{N$_V$} & \colhead{N$_I$}& \colhead{N$_V$} & \colhead{N$_I$}& \colhead{N$_V$} & \colhead{N$_I$}
}
\startdata
  16.38 & $ ~24 \pm  ~10$ & $ ~12 \pm  ~~7 $ & $ ~12 \pm   ~7$ & $ ~~8 \pm  ~~6 $ & $ ~16 \pm  ~~8$ & $~~12 \pm  ~~8 $ \nl  
  16.62 & $ ~47 \pm  ~13$ & $ ~40 \pm  ~13 $ & $ ~20 \pm   ~9$ & $ ~~4 \pm  ~~4 $ & $ ~24 \pm  ~10$ & $~~12 \pm  ~~8 $ \nl    
  16.88 & $ ~67 \pm  ~16$ & $ ~47 \pm  ~13 $ & $ ~24 \pm   10$ & $ ~20 \pm  ~~9 $ & $ ~40 \pm  ~13$ & $~~24 \pm  ~10 $ \nl    
  17.12 & $ ~74 \pm  ~16$ & $ ~73 \pm  ~16 $ & $ ~47 \pm   13$ & $ ~39 \pm  ~12 $ & $ ~20 \pm  ~11$ & $~~28 \pm  ~12 $ \nl    
  17.38 & $ ~73 \pm  ~16$ & $ ~73 \pm  ~16 $ & $ ~53 \pm   15$ & $ ~39 \pm  ~12 $ & $ ~56 \pm  ~16$ & $~~44 \pm  ~14 $ \nl    
  17.62 & $ ~94 \pm  ~18$ & $ ~87 \pm  ~18 $ & $ ~58 \pm   15$ & $ ~64 \pm  ~16 $ & $ ~84 \pm  ~18$ & $~~56 \pm  ~16 $ \nl    
  17.88 & $ 142 \pm  ~23$ & $ ~95 \pm  ~18 $ & $ ~69 \pm   17$ & $ ~62 \pm  ~15 $ & $ ~88 \pm  ~19$ & $~~83 \pm  ~19 $ \nl    
  18.12 & $ 154 \pm  ~24$ & $ 155 \pm  ~24 $ & $ 101 \pm   20$ & $ ~76 \pm  ~18 $ & $ ~83 \pm  ~19$ & $~~88 \pm  ~19 $ \nl    
  18.38 & $ 146 \pm  ~23$ & $ 155 \pm  ~24 $ & $ ~94 \pm   19$ & $ ~99 \pm  ~19 $ & $ ~88 \pm  ~20$ & $~~79 \pm  ~19 $ \nl    
  18.62 & $ 230 \pm  ~30$ & $ 182 \pm  ~26 $ & $ 101 \pm   20$ & $ ~93 \pm  ~19 $ & $ 126 \pm  ~24$ & $~~97 \pm  ~22 $ \nl    
  18.88 & $ 279 \pm  ~34$ & $ 237 \pm  ~30 $ & $ 175 \pm   26$ & $ 132 \pm  ~23 $ & $ 130 \pm  ~25$ & $~122 \pm  ~24 $ \nl    
  19.12 & $ 205 \pm  ~29$ & $ 272 \pm  ~33 $ & $ 129 \pm   23$ & $ 186 \pm  ~27 $ & $ 227 \pm  ~32$ & $~243 \pm  ~32 $ \nl    
  19.38 & $ 256 \pm  ~34$ & $ 220 \pm  ~30 $ & $ 177 \pm   26$ & $ 145 \pm  ~23 $ & $ 230 \pm  ~32$ & $~241 \pm  ~34 $ \nl    
  19.62 & $ 268 \pm  ~34$ & $ 335 \pm  ~38 $ & $ 193 \pm   28$ & $ 262 \pm  ~32 $ & $ 154 \pm  ~29$ & $~154 \pm  ~30 $ \nl    
  19.88 & $ 333 \pm  ~39$ & $ 345 \pm  ~38 $ & $ 231 \pm   31$ & $ 191 \pm  ~27 $ & $ 300 \pm  ~39$ & $~259 \pm  ~39 $ \nl    
  20.12 & $ 293 \pm  ~36$ & $ 349 \pm  ~39 $ & $ 239 \pm   31$ & $ 277 \pm  ~34 $ & $ 254 \pm  ~38$ & $~340 \pm  ~46 $ \nl    
  20.38 & $ 282 \pm  ~37$ & $ 347 \pm  ~39 $ & $ 319 \pm   38$ & $ 336 \pm  ~38 $ & $ 320 \pm  ~43$ & $~416 \pm  ~49 $ \nl    
  20.62 & $ 345 \pm  ~40$ & $ 434 \pm  ~46 $ & $ 243 \pm   33$ & $ 321 \pm  ~38 $ & $ 274 \pm  ~42$ & $~423 \pm  ~53 $ \nl    
  20.88 & $ 326 \pm  ~41$ & $ 520 \pm  ~53 $ & $ 208 \pm   31$ & $ 311 \pm  ~38 $ & $ 403 \pm  ~49$ & $~456 \pm  ~60 $ \nl    
  21.12 & $ 317 \pm  ~39$ & $ 395 \pm  ~48 $ & $ 194 \pm   30$ & $ 286 \pm  ~38 $ & $ 262 \pm  ~46$ & $~376 \pm  ~60 $ \nl    
  21.38 & $ 271 \pm  ~38$ & $ 441 \pm  ~52 $ & $ 193 \pm   30$ & $ 326 \pm  ~41 $ & $ 305 \pm  ~50$ & $~381 \pm  ~65 $ \nl    
  21.62 & $ 218 \pm  ~35$ & $ 312 \pm  ~42 $ & $ 209 \pm   31$ & $ 338 \pm  ~41 $ & $ 275 \pm  ~55$ & $~506 \pm  ~65 $ \nl    
  21.88 & $ 356 \pm  ~46$ & $ 384 \pm  ~50 $ & $ 232 \pm   35$ & $ 242 \pm  ~35 $ & $ 342 \pm  ~61$ & $~480 \pm  ~67 $ \nl    
  22.12 & $ 283 \pm  ~40$ & $ 356 \pm  ~51 $ & $ 259 \pm   35$ & $ 394 \pm  ~45 $ & $ 430 \pm  ~60$ & $~416 \pm  ~70 $ \nl    
  22.38 & $ 311 \pm  ~43$ & $ 386 \pm  ~58 $ & $ 247 \pm   35$ & $ 357 \pm  ~46 $ & $ 355 \pm  ~61$ & $~426 \pm  ~79 $ \nl    
  22.62 & $ 255 \pm  ~42$ & $ 411 \pm  ~60 $ & $ 287 \pm   39$ & $ 253 \pm  ~39 $ & $ 384 \pm  ~66$ & $~553 \pm  ~82 $ \nl    
  22.88 & $ 308 \pm  ~50$ & $ 291 \pm  ~55 $ & $ 294 \pm   40$ & $ 374 \pm  ~48 $ & $ 402 \pm  ~73$ & $~376 \pm  ~84 $ \nl    
  23.12 & $ 353 \pm  ~52$ & $ 373 \pm  ~63 $ & $ 232 \pm   37$ & $ 376 \pm  ~54 $ & $ 454 \pm  ~75$ & $~582 \pm  ~88 $ \nl    
  23.38 & $ 342 \pm  ~56$ & $ 415 \pm  ~81 $ & $ 304 \pm   43$ & $ 346 \pm  ~55 $ & $ 393 \pm  ~77$ & $~546 \pm  ~97 $ \nl    
  23.62 & $ 265 \pm  ~50$ & $ 315 \pm  ~73 $ & $ 298 \pm   44$ & $ 353 \pm  ~57 $ & $ 278 \pm  ~79$ & $~417 \pm  ~93 $ \nl    
  23.88 & $ 355 \pm  ~62$ & $ 348 \pm  102 $ & $ 330 \pm   49$ & $ 358 \pm  ~72 $ & $ 393 \pm  ~80$ & $~803 \pm  ~15 $ \nl    
  24.12 & $ 349 \pm  ~69$ & $ 198 \pm  ~85 $ & $ 232 \pm   42$ & $ 284 \pm  ~71 $ & $ 430 \pm  ~85$ & $~734 \pm  110 $ \nl    
  24.38 & $ 164 \pm  ~52$ & $ 837 \pm  289 $ & $ 256 \pm   47$ & $ 406 \pm  139 $ & $ 399 \pm  ~88$ & $~927 \pm  130 $ \nl    
  24.62 & $ 297 \pm  ~79$ &  \nodata         & $ 194 \pm   46$ & \nodata          & $ 453 \pm  ~90$ & $1126 \pm  172 $ \nl             
  24.88 & $ 237 \pm  ~89$ &  \nodata         & $ 404 \pm   83$ & \nodata          & $ 306 \pm  ~97$ & $2051 \pm  347 $ \nl      
  25.12 & $ 359 \pm  132$ &  \nodata         & $ 238 \pm   66$ & \nodata          & $ 322 \pm  ~94$ & \nodata          \nl
  25.38 & $ 732 \pm  272$ &  \nodata         & $ 247 \pm   91$ & \nodata          & $ 309 \pm  ~92$ & \nodata          \nl
  25.62 & \nodata         &  \nodata         & \nodata         & \nodata          & $ 276 \pm  110$ & \nodata          \nl
  25.88 & \nodata         &  \nodata         & \nodata         & \nodata	  & $ 633 \pm  163$ & \nodata          \nl
\enddata 
\tablenotetext{1}{Magnitudes are dereddened assuming E(B--V)=0.1}
\label{tablelf}
\end{deluxetable}

\setcounter{table}{9}

\begin{deluxetable}{llclc}
\tablewidth{0pt}
\tablecaption{Power-law Indicies}
\tablehead{
\colhead{Region} & \colhead{$x_V$} & \colhead{$\chi^2_\nu$} & \colhead{$x_I$} & \colhead{$\chi^2_\nu$} 
}
\startdata
Inner        & $0.80 \pm 0.40$        & 1.37 & $0.80 \pm 0.35$        & 1.32 \nl
Intermediate & $1.10^{+0.45}_{-0.40}$ & 0.92 & $1.15^{+0.40}_{-0.35}$ & 1.66 \nl 
Outer        & $1.30\pm{0.45}$        & 1.53 & $1.50\pm{0.40}$        & 1.83 \nl
\enddata
\label{tablex}
\end{deluxetable}

\setcounter{table}{10}

\begin{deluxetable}{ccccrcccr}
\tablewidth{0pt}
\tablecaption{$\chi^2$ Test Results.}
\tablehead{
\colhead{ } & \multicolumn{4}{c}{$V$-band} &
              \multicolumn{4}{c}{$I$-band} \nl
\colhead{Samples} &
  \colhead{ } \nl
\colhead{Compared} &
  \colhead{$V_{lim}$} & \colhead{$\chi^2_\nu$} & \colhead{$\nu$} & \colhead{P($\chi^2$)} &
  \colhead{$I_{lim}$} & \colhead{$\chi^2_\nu$} & \colhead{$\nu$} & \colhead{P($\chi^2$)}
}
\startdata
inner         & 25 & 1.34 & 34 & $0.09$ & 24 & 1.60 & 30 & $0.02$ \nl
{\it vs.}     & 24 & 1.20 & 30 & $0.21$ & 23 & 1.62 & 26 & $0.02$ \nl
intermed.     & 23 & 1.17 & 26 & $0.25$ & 22 & 1.22 & 22 & $0.21$ \nl
              & 22 & 0.96 & 22 & $0.52$ & 21 & 1.10 & 18 & $0.34$ \nl
\tablevspace{4pt}
inner         & 25 & 2.16 & 34 & $<0.01$ & 24 & 2.75 & 30 & $<0.01$ \nl
{\it vs.}     & 24 & 2.10 & 30 & $<0.01$ & 23 & 2.42 & 26 & $<0.01$ \nl
outer         & 23 & 2.26 & 26 & $<0.01$ & 22 & 2.46 & 22 & $<0.01$ \nl
              & 22 & 1.88 & 22 &  $0.01$ & 21 & 2.01 & 18 & $<0.01$ \nl
\tablevspace{4pt}
intermed.     & 25 & 1.26 & 34 & $0.15$ & 24 & 1.40 & 30 & $0.07$ \nl
{\it vs.}     & 24 & 1.16 & 30 & $0.21$ & 23 & 1.34 & 26 & $0.11$ \nl
outer         & 23 & 1.16 & 26 & $0.25$ & 22 & 1.24 & 22 & $0.20$ \nl
              & 22 & 1.12 & 22 & $0.32$ & 21 & 1.12 & 18 & $0.32$ \nl
\enddata
\label{tablechi}
\end{deluxetable}

\begin{table}
\dummytable\label{kstable}
\end{table}

\plotone{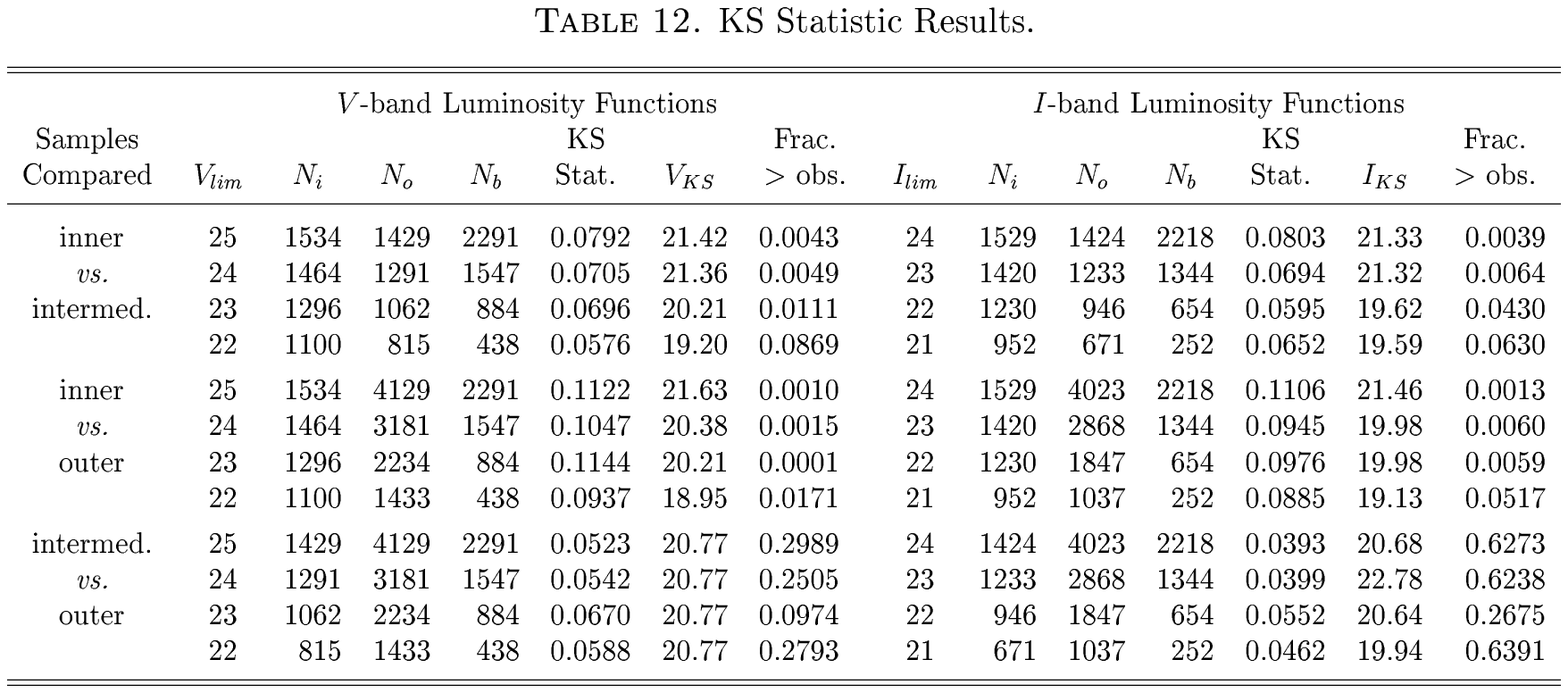}

\setcounter{table}{12}

\begin{deluxetable}{rccc}
\tablewidth{0pt}
\tablecaption{$V$-band Surface Brightness Profile.}
\tablehead{
\colhead{$\log(r)$} & \colhead{SB\tablenotemark{1}} & \colhead{$\log(r)$} & \colhead{SB} \nl
\colhead{ } & \colhead{(mag} & \colhead{ } & \colhead{(mag}  \nl
\colhead{(arcsec)} & \colhead{arcsec$^{-2}$)} & 
\colhead{(arcsec)} & \colhead{arcsec$^{-2}$)}  
}
\startdata
--0.137 & $17.13 \pm 0.06$ & 2.177 & $20.16 \pm 0.01$ \nl
  0.131 & $17.15 \pm 0.09$ & 2.231 & $20.17 \pm 0.02$ \nl
  0.254 & $17.23 \pm 0.09$ & 2.279 & $20.15 \pm 0.01$ \nl
  0.353 & $17.29 \pm 0.09$ & 2.323 & $20.17 \pm 0.01$ \nl
  0.450 & $17.37 \pm 0.10$ & 2.362 & $20.17 \pm 0.02$ \nl
  0.551 & $17.36 \pm 0.10$ & 2.398 & $20.16 \pm 0.04$ \nl
  0.654 & $17.34 \pm 0.16$ & 2.432 & $20.16 \pm 0.01$ \nl
  0.757 & $17.33 \pm 0.27$ & 2.463 & $20.16 \pm 0.01$ \nl
  0.857 & $17.69 \pm 0.37$ & 2.492 & $20.16 \pm 0.02$ \nl
  0.957 & $17.85 \pm 0.22$ & 2.519 & $20.15 \pm 0.02$ \nl
  1.057 & $18.01 \pm 0.32$ & 2.545 & $20.18 \pm 0.02$ \nl
  1.156 & $18.16 \pm 0.12$ & 2.569 & $20.16 \pm 0.14$ \nl
  1.256 & $18.73 \pm 0.16$ & 2.591 & $20.17 \pm 0.07$ \nl
  1.357 & $19.16 \pm 0.11$ & 2.613 & $20.13 \pm 0.07$ \nl
  1.457 & $19.69 \pm 0.09$ & 2.633 & $19.99 \pm 0.08$ \nl
  1.557 & $19.82 \pm 0.07$ & 2.653 & $20.05 \pm 0.09$ \nl
  1.657 & $20.02 \pm 0.04$ & 2.672 & $20.14 \pm 0.20$ \nl
  1.757 & $20.07 \pm 0.01$ & 2.690 & $20.16 \pm 0.28$ \nl
  1.857 & $20.10 \pm 0.03$ & 2.708 & $19.97 \pm 0.24$ \nl
  1.955 & $20.10 \pm 0.07$ & 2.724 & $20.17 \pm 0.38$ \nl
  2.042 & $20.16 \pm 0.06$ & 2.739 & $20.14 \pm 0.45$ \nl
  2.114 & $20.17 \pm 0.04$ &  & \nl
\enddata
\tablenotetext{1}{Background level is $V = 20.16 \pm 0.005$}
\label{sbtable}
\end {deluxetable}

\end{document}